\documentclass[twoside,english,aps,pra,twocolumn]{revtex4}
\pdfoutput=1
\usepackage{times}
\usepackage[T1]{fontenc}
\usepackage[latin1]{inputenc}
\usepackage{amsmath}
\usepackage{graphicx}
\usepackage{amssymb}

\makeatletter
\renewcommand{\vec}[1]{\mathbf{#1}}

\newcommand{\epstens}{\boldsymbol{\varepsilon}}
\newcommand{\mutens}{\boldsymbol{\mu}}
\newcommand{\p}{\tau}
\newcommand{\ptens}{\boldsymbol{\tau}}
\newcommand{\Jac}{\mathcal{J}}
\newcommand{\Jactens}{\boldsymbol{\Jac}}
\newcommand{\ftens}{\boldsymbol{\mathcal{F}}}

\usepackage{babel}
\makeatother
\begin{document}

\preprint{Preprint, submitted for publication.}

\title{Perturbation theory for anisotropic dielectric interfaces, \\
and application to sub-pixel smoothing of discretized numerical methods}

\author{Chris Kottke}

\affiliation{Research Laboratory of Electronics and Department of Mathematics,
Massachusetts Institute of Technology, Cambridge MA 02139}

\author{Ardavan Farjadpour}

\affiliation{Research Laboratory of Electronics and Department of Mathematics,
Massachusetts Institute of Technology, Cambridge MA 02139}

\author{Steven G. Johnson}

\email{stevenj@math.mit.edu}

\affiliation{Research Laboratory of Electronics and Department of Mathematics,
Massachusetts Institute of Technology, Cambridge MA 02139}

\begin{abstract}
We derive a correct first-order perturbation theory in electromagnetism
for cases where an interface between two anisotropic dielectric materials
is slightly shifted. Most previous perturbative methods give incorrect
results for this case, even to lowest order, because of the complicated
discontinuous boundary conditions on the electric field at such an
interface. Our final expression is simply a surface integral, over
the material interface, of the continuous field components from the
unperturbed structure. The derivation is based on a {}``localized''
coordinate-transformation technique, which avoids both the problem
of field discontinuities and the challenge of constructing an explicit
coordinate transformation by taking a limit in which a coordinate
perturbation is infinitesimally localized around the boundary. Not
only is our result potentially useful in evaluating boundary perturbations,
e.g. from fabrication imperfections, in highly anisotropic media such
as many metamaterials, but it also has a direct application in numerical
electromagnetism. In particular, we show how it leads to a sub-pixel
smoothing scheme to ameliorate staircasing effects in discretized
simulations of anisotropic media, in such a way as to greatly reduce
the numerical errors compared to other proposed smoothing schemes. 
\end{abstract}
\maketitle

\section{introduction}

In this paper, we present a technique to apply perturbative techniques
to Maxwell's equations with anisotropic materials, in particular for
the case where the position of an interface between two such materials
is perturbed, generalizing an earlier result for isotropic materials
\cite{JohnsonIb02}. In this case, the discontinuities of the fields
at the interface cause many standard perturbative methods to fail,
which is unfortunate because such methods are very useful for many
problems in electromagnetism where one wishes to study the effect
of small deviations from a given structure---not only do perturbative
methods allow one to apply the computational efficiency of idealized
problems to more realistic situations, but they may also offer greater
analytical insight than brute-force numerical approaches. The corrected
solution described in this paper should aid the study of interface
perturbations, from surface roughness to fiber birefringence, in the
context of anisotropic materials. Such materials have become increasingly
important thanks to the discovery of {}``metamaterials,'' subwavelength
composite structures that simulate homogeneous media with unusual
properties such as negative refractive indices \cite{Smith04}, and
which may be strongly anisotropic in certain applications---for example,
those involving spherical or cylindrical geometries \cite{Pendry03}
such as recent proposals for {}``invisibility'' cloaks \cite{Pendry06}.
Furthermore, we have recently shown that interface-perturbation analyses
benefit even purely brute-force computations, because they enable
the design of sub-pixel smoothing techniques that greatly increase
the accuracy (and may even increase the \emph{order} of convergence)
of discretized methods \cite{Farjadpour06}, which are normally degraded
by discontinuous interfaces \cite{JohnsonJo01,Ditkowski01}. Here,
we show that our corrected perturbation analysis provides similar
benefits for modeling anisotropic materials, where it yields a second-order
accurate smoothing technique (correcting a \emph{}previous heuristic
proposal \cite{JohnsonJo01}). 

There have been several previous approaches to rigorous treatment
of interface perturbations in electromagnetism, where classic approaches
for small $\Delta\varepsilon$ perturbations fail because of the field
discontinuities \cite{Hill81,Lohmeyer99,JohnsonIb02,Skorobogatiy02:curvi,Skorobogatiy04,Johnson05:bump}.
One approach that was applied successfully to boundaries between isotropic
materials is essentially to guess the correct form of the perturbation
integral and then to prove \emph{a posteriori} that it is correct
\cite{JohnsonIb02}. For isotropic materials, where there is some
guidance from effective-medium heuristics \cite{Meade93}, this was
practical, but the correct answer (below) appears to be much more
difficult to guess for anisotropic materials. Another approach, which
generalizes to the more difficult case of small surface {}``bumps''
that are not locally flat, was to express the problem in terms of
finding the polarizability of the perturbation and then connecting
it back to the perturbation integral via the method of images \cite{Johnson05:bump}.
For a locally flat perturbation between isotropic materials, this
process can be carried out analytically to reproduce the previous
result from Ref\@.~\cite{JohnsonIb02}, but it becomes rather complicated
for anisotropic media. Third, one can transform the problem into a
statement about the coordinate system to avoid problems of shifting
field discontinuities, by finding a coordinate transformation that
expresses the interface shift \cite{Skorobogatiy02:curvi,Skorobogatiy04}.
This approach, while powerful, has two shortcomings: first, finding
an explicit coordinate transformation may be difficult for a complicated
interface perturbation; and second, the resulting perturbation integrals
are expressed in terms of the fields everywhere in space, not just
at the boundaries. Intuitively, one expects that the effect of the
perturbation should depend only on the field at the boundaries, as
was found explicitly for the isotropic case \cite{JohnsonIb02,Johnson05:bump}.
In this paper, we derive precisely such an expression for the case
of interfaces between anisotropic materials, by developing a general
new analytical technique for interface perturbations: we express the
perturbation as a coordinate transformation, but using a coordinate
transform \emph{localized} around the perturbed interface, and take
a limit in which this localization becomes narrower and narrower so
that the choice of transform disappears from the final result.

In the following sections, we first formulate the problem of the effect
of an interface perturbation more precisely, relate our formulation
to other possibilities, and summarize our final result in the form
of eq\@.~(\ref{eq:pert-surf}). We then derive quite generally how
to formulate the problem of interface perturbations in terms of a
localized coordinate transformation, and show how this allows us to
express the perturbation-theory integral as a sum of contributions
around individual \emph{points} on the interface. Next, we apply this
framework to the specific problem of a boundary between two anisotropic
dielectric materials, and derive our final result. As a check, our
perturbation theory is then validated against brute-force computations
for a simple numerical example. Finally, we discuss the application
of our new perturbation result to sub-pixel smoothing of discretized
numerical methods, and show that we obtain a smoothing technique that
leads to much more accurate results at a given spatial resolution.
In the appendix, we provide a compact derivation and generalization
of a useful result \cite{Ward96} relating coordinate transformations
to changes in $\varepsilon$ and $\mu$.

\section{Problem Formulation\label{sec:Problem-Formulation}}

There are many ways to formulate perturbation techniques in electromagnetism.
One common formulation, analogous to {}``time-independent perturbation
theory'' in quantum mechanics \cite{Tannoudji77}, is to express
Maxwell's equations as a generalized Hermitian eigenproblem $\nabla\times\nabla\times\vec{E}=\omega^{2}\varepsilon\vec{E}$
in the frequency $\omega$ and electric field $\vec{E}$ (or equivalent
formulations in terms of the magnetic field $\vec{H}$) \cite{Joannopoulos95},
and then to consider the first-order change $\Delta\omega$ in the
frequency from a small change $\Delta\varepsilon$ in the dielectric
function $\varepsilon(\vec{x})$ (assumed real and positive), which
turns out to be \cite{Joannopoulos95}: \begin{equation}
\frac{\Delta\omega}{\omega}=-\frac{\int\vec{E}^{*}\cdot\Delta\varepsilon\vec{E}\, d^{3}\vec{x}}{2\int\vec{E}^{*}\cdot\varepsilon\vec{E}\, d^{3}\vec{x}}+O(\Delta\varepsilon^{2}),\label{eq:freq-pert}\end{equation}
where $\vec{E}$ and $\omega$ are the electric field and eigenfrequency
of the \emph{unperturbed} structure $\varepsilon$, respectively,
and $*$ denotes complex conjugation. The key part of this expression
is the numerator of the right-hand side, which is what expresses the
effect of the perturbation, and this same numerator appears in a nearly
identical form for many different perturbation techniques. For example,
one obtains a similar expression in: finding the perturbation $\Delta\beta$
in the propagation constant $\beta$ of a waveguide mode \cite{Johnson01:og};
the coupling coefficient ($\sim\int\vec{E}^{*}\cdot\Delta\varepsilon\vec{E}'$)
between two modes $\vec{E}$ and $\vec{E}'$ in coupled-wave theory
\cite{Marcuse74,Katsenelenbaum98,Johnson02:adiabatic}; or the scattering
current $\vec{J}\sim\Delta\varepsilon\vec{E}$ (and the scattered
power $\sim\int\vec{J}^{*}\cdot\vec{E}$) in the {}``volume-current''
method (equivalent to the first Born approximation) \cite{Snyder83,Kuznetsov83,Chew95,Johnson05:bump}.
Eq\@.~(\ref{eq:freq-pert}) also corresponds to an \emph{exact}
result for the \emph{derivative} of $\omega$ with respect to any
parameter $p$ of $\varepsilon$, since if we write $\Delta\varepsilon=\frac{\partial\varepsilon}{\partial p}\Delta p+O(\Delta p^{2})$
we can divide both sides by $\Delta p$ and take the limit $\Delta p\rightarrow0$;
this result is equivalent to the Hellman-Feynman theorem of quantum
mechanics \cite{Tannoudji77,JohnsonIb02}. In cases where the unperturbed
$\varepsilon$ is not real, corresponding to absorption or gain, or
when one is considering {}``leaky modes,'' the eigenproblem typically
becomes complex-symmetric rather than Hermitian and one obtains a
similar formula but without the complex conjugation \cite{LeungLi94}.
Therefore, any modification to the form of this numerator for the
frequency-perturbation theory immediately leads to corresponding modified
formulas in many other perturbative techniques, and it is sufficient
for our purposes to consider frequency-perturbation theory only.%
\begin{figure}
\begin{centering}\includegraphics[width=1\columnwidth]{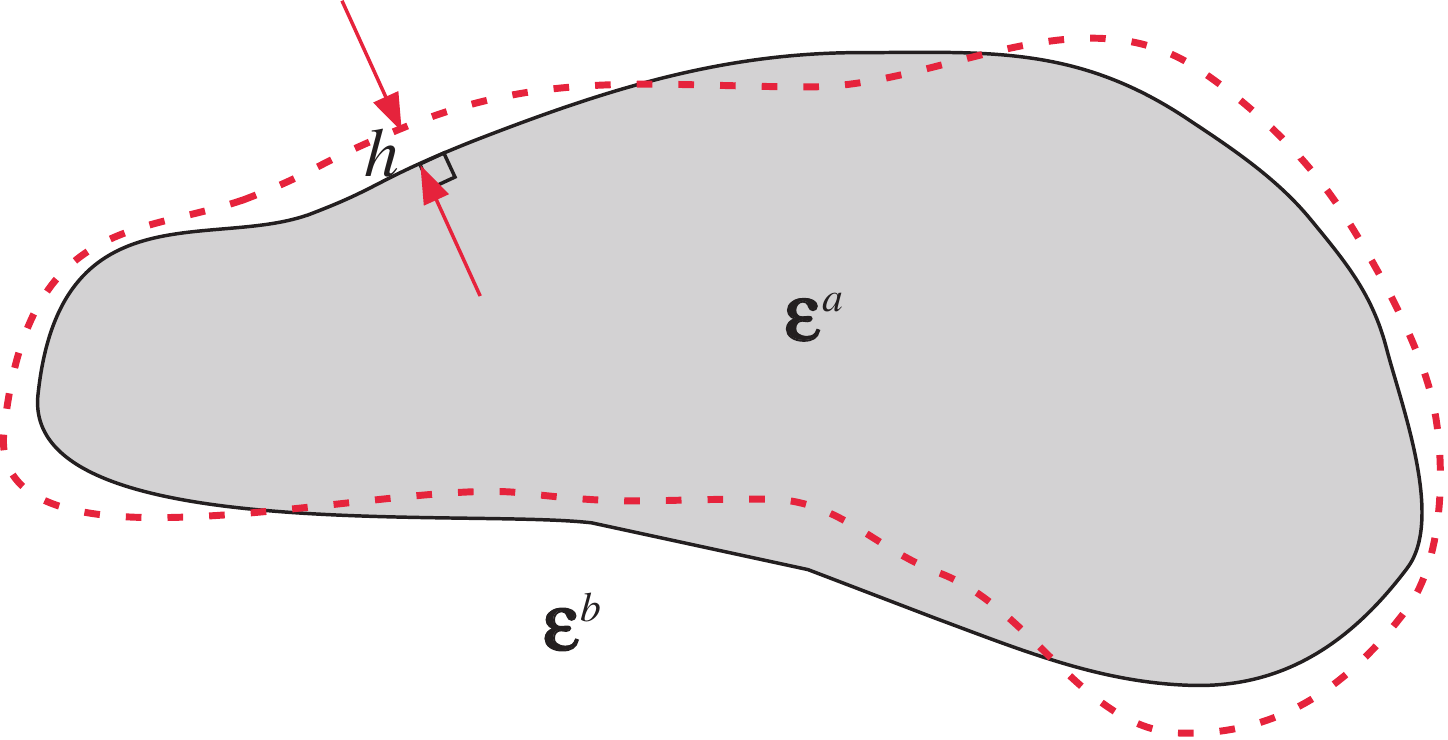}\par\end{centering}

\caption{\label{fig:surface-pert}Schematic of an interface perturbation:
the interface between two materials $\epstens^{a}$ and $\epstens^{b}$
(possibly anisotropic) is shifted by some small position-dependent
displacement $h$.}
\end{figure}

As we showed in Ref\@.~\cite{JohnsonIb02}, eq\@.~(\ref{eq:freq-pert})
is not valid when $\Delta\varepsilon$ is due to a small change in
the position of a boundary between two dielectric materials (except
in the limit of low dielectric contrast), but a simple correction
is possible. In particular, let us consider situations like the one
shown in Fig\@.~\ref{fig:surface-pert}, where the dielectric boundary
between two materials $\varepsilon^{a}$ and $\varepsilon^{b}$ is
shifted by some small displacement $h$ (which may be a function of
position). Directly applying eq\@.~(\ref{eq:freq-pert}), with $\Delta\varepsilon=\pm(\varepsilon^{a}-\varepsilon^{b})$
in the regions where the material has changed, gives an incorrect
result, and in particular $\Delta\omega/h$ (which should ideally
go to the exact derivative $d\omega/dh$) is incorrect even for $h\rightarrow0$.
The problem turns out to be not so much that $\Delta\varepsilon$
is not small, but rather that $\vec{E}$ is discontinuous at the boundary,
and the standard method in the limit $h\rightarrow0$ leads to an
ill-defined surface integral of $\vec{E}$ over the interfaces. For
\emph{isotropic} materials, corresponding to \emph{scalar} $\varepsilon^{a,b}$,
the correct numerator turns out to be, instead, the following surface
integral over the boundary \cite{JohnsonIb02}: \begin{multline}
\int\vec{E}^{*}\cdot\Delta\varepsilon\vec{E}\, d^{3}\vec{x}\longrightarrow\\
\iint\left[\left(\varepsilon^{a}-\varepsilon^{b}\right)\left|\vec{E}_{\Vert}\right|^{2}-\left(\frac{1}{\varepsilon^{a}}-\frac{1}{\varepsilon^{b}}\right)\left|D_{\perp}\right|^{2}\right]\vec{h}\cdot d\vec{A},\label{eq:pert-surf-scalar}\end{multline}
where $\vec{E}_{\Vert}$ and $D_{\perp}$ are the (continuous) components
of $\vec{E}$ and $\vec{D}=\varepsilon\vec{E}$ parallel and perpendicular
to the boundary, respectively, $d\vec{A}$ points towards $\varepsilon^{b}$,
and $\vec{h}$ is the displacement of the interface from $\varepsilon^{a}$
towards $\varepsilon^{b}$. 

In this paper, we will generalize eq\@.~(\ref{eq:pert-surf-scalar})
to handle the case where the two materials are \emph{anisotropic},
corresponding to arbitrary $3\times3$ tensors $\epstens^{a}$ and
$\epstens^{b}$ (assumed Hermitian and positive-definite to obtain
a well-behaved Hermitian eigenproblem). In the generalized case, it
is convenient to define a local coordinate frame $(x_{1},x_{2},x_{3})$
at each point on the surface, where the $x_{1}$ direction is orthogonal
to the surface and the $(x_{2},x_{3})$ directions are parallel. We
also define a continuous field {}``vector'' $\vec{F}=(D_{1},E_{2},E_{3})$
so that $F_{1}=D_{\perp}$ and $\vec{F}_{2,3}=\vec{E}_{\Vert}$. As
derived below, the resulting numerator of eq\@.~(\ref{eq:freq-pert}),
generalizing eq\@.~(\ref{eq:pert-surf-scalar}), is: \begin{equation}
\iint\vec{F}^{*}\cdot\left[\ptens\left(\epstens^{a}\right)-\ptens\left(\epstens^{b}\right)\right]\cdot\vec{F}\,\vec{h}\cdot d\vec{A},\label{eq:pert-surf}\end{equation}
where $\ptens(\epstens)$ is the $3\times3$ matrix: \begin{equation}
\ptens(\epstens)=\left(\begin{array}{ccc}
-\frac{1}{\varepsilon_{11}} & \frac{\varepsilon_{12}}{\varepsilon_{11}} & \frac{\varepsilon_{13}}{\varepsilon_{11}}\\
\frac{\varepsilon_{21}}{\varepsilon_{11}} & \quad\varepsilon_{22}-\frac{\varepsilon_{21}\varepsilon_{12}}{\varepsilon_{11}}\quad & \varepsilon_{23}-\frac{\varepsilon_{21}\varepsilon_{13}}{\varepsilon_{11}}\\
\frac{\varepsilon_{31}}{\varepsilon_{11}} & \varepsilon_{32}-\frac{\varepsilon_{31}\varepsilon_{12}}{\varepsilon_{11}} & \varepsilon_{33}-\frac{\varepsilon_{31}\varepsilon_{13}}{\varepsilon_{11}}\end{array}\right),\label{eq:ptens-definition}\end{equation}
which reduces to eq\@.~(\ref{eq:pert-surf-scalar}) when $\epstens$
is a scalar multiple $\varepsilon$ of the identity matrix. (Our assumption
that $\epstens$ is positive-definite guarantees that $\varepsilon_{11}>0$.)

We should note an important restriction: eq\@.~(\ref{eq:pert-surf-scalar})
and eq\@.~(\ref{eq:pert-surf}) require that the radius of curvature
of the interface be much larger than $h=|\vec{h}|$, except possibly
on a set of measure zero (such as at isolated corners or edges). Otherwise,
more complicated methods must be employed \cite{Johnson05:bump}.
For example one cannot apply the above equations to the case of a
hemispherical {}``bump'' of radius $h$ on the unperturbed surface,
in which case the lowest order perturbation is $\Delta\omega\sim O(h^{3})$
and requires a small numerical computation of the polarizability of
the hemisphere \cite{Johnson05:bump}.

\section{Local Coordinate Perturbations}

The difficulty with applying the standard perturbation-theory result
(\ref{eq:freq-pert}) to a boundary perturbation is that, instead
of a \emph{small} $\Delta\varepsilon$ with \emph{fixed} boundary
conditions on the fields (to lowest order), we have a \emph{large}
$\Delta\varepsilon$ over a small region in which the field boundary
discontinuities have \emph{shifted}. However, we can transform one
problem into the other: we construct a coordinate transformation that
maps the new boundary location back onto the old boundary, so that
in the new coordinates the boundary conditions are unaltered while
there is a small change in the differential operators due to the coordinate
shift. In expressing the problem in this fashion, we will present
two key techniques. First, we employ a result from \cite{Ward96},
generalized in the Appendix to anisotropic materials, that expresses
an arbitrary coordinate transform as a change $\Delta\epstens$ and
$\Delta\mutens$ in the permittivity and permeability tensors, which
allows us to directly apply eq\@.~(\ref{eq:freq-pert}). Second,
unlike Refs\@.~\cite{Skorobogatiy02:curvi,Skorobogatiy04}, we do
not wish to \emph{explicitly} construct any coordinate transformation,
since this may become very complicated for an arbitrary perturbation
in an arbitrary-shaped boundary. Instead, we express the boundary
shift in terms of a \emph{local} coordinate transform, that only {}``nudges''
the coordinates near the perturbed boundary, and in the limit where
the region of this coordinate perturbation becomes arbitrarily small
we will recover the coordinate-independent surface integrals (\ref{eq:pert-surf-scalar})
and (\ref{eq:pert-surf}).

\subsection{Coordinate perturbations\label{sub:Coordinate-perturbations}}

Suppose that in a certain coordinate system $\vec{x}$ we have electric
field $\vec{E}(\vec{x},t)$, magnetic field $\vec{H}(\vec{x},t)$,
dielectric tensor $\epstens(\vec{x})$, and relative magnetic permeability
tensor $\mutens(\vec{x})$, satisfying the Euclidean Maxwell's equations.
Now, we transform to some new coordinates $\vec{x}'(\vec{x})$, with
a $3\times3$ Jacobian matrix $\Jactens$ defined by $\Jac_{ij}=\frac{\partial x_{i}'}{\partial x_{j}}$.
In the new coordinates, the fields can still be written as the solution
of the Euclidean Maxwell's equations if the following transformations
are made in addition to the change of coordinates: \begin{equation}
\vec{E}'=(\Jactens^{T})^{-1}\cdot\vec{E},\label{eq:E-trans}\end{equation}
 \begin{equation}
\vec{H}'=(\Jactens^{T})^{-1}\cdot\vec{H},\label{eq:H-trans}\end{equation}
\begin{equation}
\epstens'=\frac{\Jactens\cdot\epstens\cdot\Jactens^{T}}{\det\Jactens},\label{eq:eps-trans}\end{equation}
\begin{equation}
\mutens'=\frac{\Jactens\cdot\mutens\cdot\Jactens^{T}}{\det\Jactens},\label{eq:mu-trans}\end{equation}
 where $\Jactens^{T}$ denotes the transpose. This result is derived
in the Appendix, generalized from the result for scalar $\varepsilon$
and $\mu$ from Ref\@.~\cite{Ward96}.

Now, suppose the coordinate change is {}``small,'' meaning that
$\Jactens=\vec{1}+\Delta\Jactens$, where the eigenvalues of $\Delta\Jactens(\vec{x})$
are everywhere $O(\delta)$ for some small parameter $\delta$. Then
$\Delta\epstens(\vec{x}')=\epstens'(\vec{x}')-\epstens[\vec{x}(\vec{x}')]=O(\delta)$
and similarly $\Delta\mutens=O(\delta)$. Therefore, the solutions
of Maxwell's equations will be \emph{nearly} those of $\epstens$
and $\mutens$ merely translated to the new coordinate locations,
and the difference due to $\Delta\epstens$ and $\Delta\mutens$ can
be accounted for, to $O(\delta^{2})$, by first-order perturbation
theory. That is, generalizing eq\@.~(\ref{eq:freq-pert}) to the
case of anisotropic media with both $\epstens$ and $\mutens$, one
finds by elementary perturbation theory for the generalized eigenproblem:\begin{align}
\frac{\Delta\omega}{\omega_{0}} & =-\frac{\int\left[\vec{E}_{0}^{*}\cdot\Delta\epstens\cdot\vec{E}_{0}+\vec{H}_{0}^{*}\cdot\Delta\mutens\cdot\vec{H}_{0}\right]\, d^{3}\vec{x}'}{\int\left[\vec{E}_{0}^{*}\cdot\epstens\cdot\vec{E}_{0}+\vec{H}_{0}^{*}\cdot\mutens\cdot\vec{H}_{0}\right]\, d^{3}\vec{x}'}+O(\delta^{2})\nonumber \\
 & =-\frac{\int\left[\vec{E}_{0}^{*}\cdot\Delta\epstens\cdot\vec{E}_{0}+\vec{H}_{0}^{*}\cdot\Delta\mutens\cdot\vec{H}_{0}\right]\, d^{3}\vec{x}'}{2\int\vec{E_{0}}^{*}\cdot\epstens\cdot\vec{E}_{0}\, d^{3}\vec{x}'}+O(\delta^{2}),\label{eq:freq-pert-aniso}\end{align}
where the {}``0'' subscripts denote the solution for the unperturbed
system, given by $\epstens[\vec{x}(\vec{x}')]$ and $\mutens[\vec{x}(\vec{x}')]$,
i.e. $\epstens$ and $\mutens$ simply translated into the $\vec{x}'$
coordinates without transforming by the Jacobian factors.

\subsection{Interface-localized coordinate transforms}

Suppose that we have an unperturbed interface between two materials
$\epstens^{a}$ and $\epstens^{b}$ that forms a surface $S_{0}$
(i.e., the points $\vec{x}_{0}\in S_{0}$), and we perturb it to a
new interface $S$ by a small perpendicular shift $\vec{h}(\vec{x})$
as depicted schematically in Fig\@.~\ref{fig:surface-pert}. In
order to investigate this boundary shift, we will perform a coordinate
transform $\vec{x}'(\vec{x})$ that shifts $S$ to $S'=S_{0}$. That
is, in our new coordinates, the interface has not been perturbed,
but the materials have changed by the Jacobian factors as described
in the previous section. Moreover, we will construct our coordinate
transform so that it is \emph{localized} to the interface, i.e. so
that $\vec{x}'=\vec{x}$ far from $S_{0}$. In particular, we write:
\begin{equation}
\vec{x}'=\vec{x}-\vec{h}(\vec{x})L(\vec{x})\label{eq:coord-shift}\end{equation}
where $L(\vec{x})\in[0,1]$ is some differentiable localized function,
equal to unity on the interface {[}$L(S)=1$] and identically zero
outside some small radius-$R/2$ neighborhood of the interface (the
\emph{support} of $L$ lies within this neighborhood), chosen so that
$|\nabla L|=O(1/R)$. Eq\@.~(\ref{eq:coord-shift}) is constructed
so that $\vec{x}\in S$ implies $\vec{x}'\in S'=S_{0}$, causing the
new interface $S$ to be mapped to $S_{0}$ as desired. Thus, $\vec{h}(\vec{x})$
for $\vec{x}\in S$ must be the perpendicular displacement from $S_{0}$
to $S$. For $\vec{x}\notin S$, $\vec{h}(\vec{x})$ should be some
differentiable, slowly-varying function (except possibly at isolated
surface kinks and discontinuities). The precise functions $L$ and
$\vec{h}$ will turn out to be irrelevant to our final answer (\ref{eq:pert-surf}),
so we need not construct them explicitly.

We will take $|\vec{h}|=h\ll1$ to be the small parameter of our perturbation
theory, and will concern ourselves with obtaining the correct first-order
$\Delta\omega$ in the limit $h\rightarrow0$. We will also eventually
take the limit $R\rightarrow0$, but will still require $h\ll R$
in order to ensure, as shall become apparent below, that the Jacobian
factor of the coordinate transformation remains close to unity. (That
is, we let $h$ go to zero faster than $R$.) Finally, in order to
have $\vec{h}(\vec{x})$ be sufficiently slowly varying that we can
neglect its derivatives compared to the derivatives of $L(\vec{x})$,
below, it will be important to require that the radius of curvature
of $S_{0}$ and $S$ be much larger than $h$, except possibly at
isolated points; otherwise, more complicated perturbative methods
are required \cite{Johnson05:bump}.

\subsection{Point-localized coordinate transforms}

The coordinate transformation (\ref{eq:coord-shift}) representing
our boundary perturbation is localized around the perturbed interface,
but is convenient to go one step further: we will represent the coordinate
transform as a \emph{summation} of coordinate transformations localized
around individual \emph{points} on the interface, by exploiting the
concept of a \emph{partition of unity} from topology \cite{Munkres75},
reviewed below.

Consider the support of the function $L(\vec{x})$ from above. This
support is covered by the open set of spherical radius-$R$ neighborhoods
of every point on the surface, and that covering must admit a locally
finite subcovering $\left\{ U^{(\alpha)}\right\} $; that is, a subset
of neighborhoods $\left\{ U^{(\alpha)}\right\} $ such that every
point on the surface intersects finitely many neighborhoods $U^{(\alpha)}$,
and the union of the $U^{(\alpha)}$ covers the support of $L$. There
must also exist a partition of unity $\left\{ \phi^{(\alpha)}\right\} $:
a set of differentiable functions $\phi^{(\alpha)}(\vec{x})\in[0,1]$
with support $\subseteq U^{(\alpha)}$, such that $\sum_{\alpha}\phi^{(\alpha)}(\vec{x})=1$
everywhere in the support of $L$. We can then write \begin{equation}
L(\vec{x})=\left[\sum_{\alpha}\phi^{(\alpha)}(\vec{x})\right]L(\vec{x})=\sum_{\alpha}K^{(\alpha)}(\vec{x}),\label{eq:L-partition}\end{equation}
where each $K^{(\alpha)}(\vec{x})=\phi^{(\alpha)}(\vec{x})\, L(\vec{x})\in[0,1]$
is a differentiable function localized to a small radius-$R$ neighborhood
$U^{(\alpha)}$ of a single point on the interface. The Jacobian $\Jactens$
of the coordinate transformation (\ref{eq:coord-shift}) can then
be written in the form: \begin{equation}
\Jactens=\vec{1}+\sum_{\alpha}\Delta\Jactens^{(\alpha)},\label{eq:Jac-partition}\end{equation}
where \begin{equation}
\Delta\Jac_{ij}^{(\alpha)}=-\frac{\partial}{\partial x_{j}}\left[h_{i}(\vec{x})\, K^{(\alpha)}(\vec{x})\right]\label{eq:Delta-Jactens-alpha}\end{equation}
has support $\subseteq U^{(\alpha)}$. 

The key advantage of this construction arises if we look at $\Delta\epstens=\epstens'-\epstens$
from eq\@.~(\ref{eq:eps-trans}). Assuming $\Delta\Jactens$ is
small and we are computing $\Delta\epstens$ to first-order, then
we can write $\Delta\epstens=\sum_{\alpha}\Delta\epstens^{(\alpha)}$
as a sum of contributions from each $\Delta\Jactens^{(\alpha)}$ individually,
and similarly for $\Delta\mutens$. Therefore, when computing the
first-order perturbation $\Delta\omega$ from eq\@.~(\ref{eq:freq-pert-aniso}),
we can write $\Delta\omega=\sum_{\alpha}\Delta\omega^{(\alpha)}$
as a sum of contributions $\Delta\omega^{(\alpha)}$ analyzed in each
point neighborhood separately. This removes the need to deal with
the complex shape of the entire boundary at once, and is the procedure
that we adopt in the following section.

\section{Perturbation theory derivation}

In the previous section, we established several important preliminary
results that allow us to express a boundary perturbation, via coordinate
transformation, as a sum of localized material perturbations $\Delta\epstens^{(\alpha)}$
and $\Delta\mutens^{(\alpha)}$ around individual points of the boundary.
We will now explicitly evaluate those contributions, taking the limit
as the perturbation $h\rightarrow0$ and the coordinate distortion
radius $R\rightarrow0$ to obtain our coordinate-independent final
result, eq.~(\ref{eq:pert-surf}).%
\begin{figure}
\begin{centering}\includegraphics[width=0.6\columnwidth]{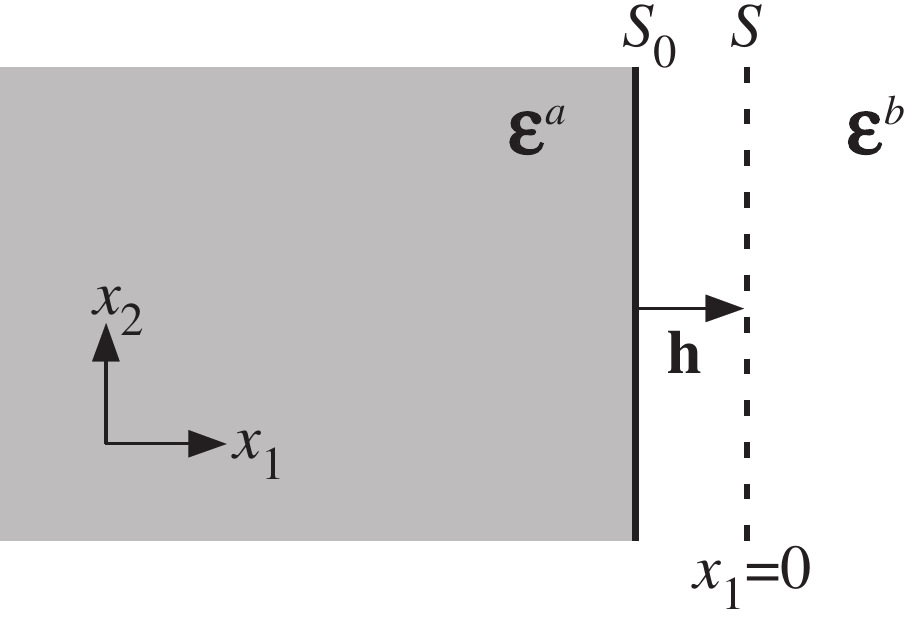}\par\end{centering}

\caption{\label{fig:surface-pert-zoom}Schematic of an interface perturbation
as in Fig\@.~\ref{fig:surface-pert}, magnifying a small portion
of the interface where the surface is locally flat. A local coordinate
frame $(x_{1},x_{2},x_{3})$ is chosen so that $x_{1}$ is perpendicular
to the surface, and so that $x_{1}=0$ denotes the location of the
perturbed surface $S$ (shifted perpendicularly by $\vec{h}$ from
the original surface $S_{0}$).}
\end{figure}

We therefore restrict our attention to a single neighborhood $U^{(\alpha)}$
and the contribution from the corresponding term $K^{(\alpha)}$ in
the coordinate transformation. In this small neighborhood of radius
$R$, we can take $\vec{h}(\vec{x})\approx\vec{h}^{(\alpha)}$ to
be a constant to lowest order in $R$. In this case, the interface
is locally flat, and we can choose a local coordinate frame $(x_{1},x_{2},x_{3})$
so that $x_{1}$ is the direction perpendicular to the interface at
$x_{1}=0$, with $x_{1}<0$ corresponding to $\epstens^{a}$ and $x_{1}>0$
corresponding to $\epstens^{b}$, as shown in Fig\@.~\ref{fig:surface-pert-zoom}.
In this coordinate frame $\vec{h}^{(\alpha)}=(h^{(\alpha)},0,0)$,
the Jacobian contribution $\Delta\Jactens^{(\alpha)}$ simplifies
to: \begin{equation}
\Delta\Jac_{ij}^{^{(\alpha)}}=-\delta_{i1}h^{(\alpha)}K_{j}^{(\alpha)}+O(h^{(\alpha)}R),\label{eq:delta-Jac-alpha-h}\end{equation}
where $\delta_{i1}$ is the Kronecker delta and $K_{j}^{(\alpha)}$
denotes $\partial K^{(\alpha)}/\partial x_{j}$. Since $R\gg|\vec{h}|$
by assumption and $K^{(\alpha)}\in[0,1]$ is a smooth localized function
with support of radius $R$, $K^{(\alpha)}$ can be constructed so
that $h^{(\alpha)}K_{j}^{(\alpha)}=O(h/R)$, i.e so that the derivatives
are small. This will make $\Jactens$ close to unity and allow us
to use the perturbation equation~(\ref{eq:freq-pert-aniso}).

We must now construct $\Delta\epstens^{(\alpha)}$ to first order.
Since $\Jactens=\mathbf{1}+\sum_{\alpha}\Delta\Jactens^{(\alpha)}$,
we obtain: \begin{equation}
\frac{1}{\det\Jactens}=1+\sum_{\alpha}h^{(\alpha)}K_{1}^{(\alpha)}+O(h^{2})+O(hR).\label{eq:det-Jac}\end{equation}
Combined with eq.~(\ref{eq:Delta-Jactens-alpha}), we can now evaluate
eq\@.~(\ref{eq:eps-trans}) for $\epstens'$, to lowest order, to
obtain $\Delta\epstens=\sum_{\alpha}\Delta\epstens^{(\alpha)}+O(h^{2})+O(hR)$,
with \begin{equation}
\Delta\varepsilon_{ij}^{(\alpha)}=\left[\varepsilon_{ij}K_{1}^{(\alpha)}-\sum_{k}K_{k}^{(\alpha)}(\delta_{i1}\varepsilon_{kj}+\delta_{j1}\varepsilon_{ik})\right]h^{(\alpha)}.\label{eq:delta-eps-alpha}\end{equation}
This will contribute to (\ref{eq:freq-pert-aniso}) via the integral:
\begin{equation}
I^{(\alpha)}=\int_{U^{(\alpha)}}\vec{E}^{*}\cdot\Delta\epstens^{(\alpha)}\cdot\vec{E}\, d^{3}\vec{x},\label{eq:I-alpha}\end{equation}
where we have dropped the {}``$0$'' subscript from the unperturbed
field $\vec{E}$ for simplicity. In order to simplify this integral,
we will write $\vec{E}=(E_{1},E_{2},E_{3})$ in terms of $\vec{F}=(D_{1},E_{2},E_{3})$,
since $\vec{F}$ is continuous whereas $E_{1}$ is not. Solving for
$E_{1}$ in $\vec{D}=\epstens\cdot\vec{E}$ yields $E_{1}=\frac{1}{\varepsilon_{11}}(D_{1}-\varepsilon_{12}E_{2}-\varepsilon_{13}E_{3})$,
and thus $\vec{E}=\ftens\cdot\vec{F}$ where \begin{equation}
\ftens(\epstens)=\left(\begin{array}{ccc}
\frac{1}{\varepsilon_{11}} & -\frac{\varepsilon_{12}}{\varepsilon_{11}} & -\frac{\varepsilon_{13}}{\varepsilon_{11}}\\
 & 1\\
 &  & 1\end{array}\right).\label{eq:ftens}\end{equation}
Because $\vec{F}$ is continuous, we can write $\vec{F}(\vec{x})=\vec{F}^{(\alpha)}+O(R)$,
where the $O(R)$ term is a higher-order contribution to $I^{(\alpha)}$
that can be dropped and the $\vec{F}^{(\alpha)}$ is a constant that
can be pulled out of the integral. Therefore, we are left with\begin{equation}
I^{(\alpha)}=\vec{F}^{(\alpha)*}\cdot\left[\int_{U^{(\alpha)}}\ftens^{\dagger}\cdot\Delta\epstens^{(\alpha)}\cdot\ftens\, d^{3}\vec{x}\right]\cdot\vec{F}^{(\alpha)}+O(hR),\label{eq:I-alpha-F}\end{equation}
where $\ftens^{\dagger}$ is the conjugate-transpose. This integral
now simplifies a great deal, because the only non-constant terms are
from the $K_{j}^{(\alpha)}$ and the step-function $\Theta(x_{1})$
dependence of $\epstens(\vec{x})$. In particular, the integrals over
the $K_{2}^{(\alpha)}$ and $K_{3}^{(\alpha)}$ terms vanish, because
along the $x_{2}$ and $x_{3}$ directions, respectively, they are
integrals of the derivatives of a function $K^{(\alpha)}$ that vanishes
at the endpoints. We are left with the $K_{1}^{(\alpha)}$ terms,
which yield the integrand: \begin{multline}
\ftens^{\dagger}\cdot\left(\begin{array}{ccc}
-\varepsilon_{11}\\
 & \varepsilon_{22} & \varepsilon_{23}\\
 & \varepsilon_{32} & \varepsilon_{33}\end{array}\right)\cdot\ftens\, K_{1}^{(\alpha)}h^{(\alpha)}=\ptens(\epstens)\, K_{1}^{(\alpha)}h^{(\alpha)}\\
=\left\{ \ptens(\epstens^{a})+\left[\ptens(\epstens^{b})-\ptens(\epstens^{a})\right]\,\Theta(x_{1})\right\} \, K_{1}^{(\alpha)}h^{(\alpha)}\label{eq:K1-integrand}\end{multline}
where the product of the three matrices gives precisely the matrix
$\ptens(\epstens)$ defined in eq\@.~(\ref{eq:ptens-definition}),
using the assumption that $\epstens$ is Hermitian ($\epstens^{\dagger}=\epstens$).
When eq\@.~(\ref{eq:K1-integrand}) is integrated by parts in the
$x_{1}$ direction, we obtain the integral of $K^{(\alpha)}$ multiplied
by a delta function $\delta(x_{1})$ from the derivative of $\Theta(x_{1})$,
producing: \begin{equation}
\iint\left[\ptens(\epstens^{a})-\ptens(\epstens^{b})\right]\, K^{(\alpha)}(0,x_{2},x_{3})h^{(\alpha)}dx_{2}dx_{3}.\label{eq:I-alpha-final}\end{equation}
When this is summed over $\alpha$ to obtain the total perturbation
integral, however, $\sum_{\alpha}K^{(\alpha)}(0,x_{2},x_{3})=L(0,x_{2},x_{3})=1$
by construction (since $L=1$ on the interface). Thus, we obtain the
surface integral of eq\@.~(\ref{eq:pert-surf}), as desired, where
$h^{(\alpha)}dx_{2}dx_{3}=\vec{h}\cdot d\vec{A}$.

The analysis of the $\Delta\mutens^{(\alpha)}$ term proceeds identically,
although here the continuous field components are $(B_{1},H_{2,}H_{3})$,
but in this case it yields zero if $\mutens^{a}=\mutens^{b}$ (as
in the common case of non-magnetic materials where $\mu$ is identically
1).

\section{Numerical Validation}

To check the correctnessf of the perturbative analysis above, we performed
the following numerical computation. We solve the full-vector Maxwell
eigenproblem numerically, for inhomogeneous anisotropic dielectric
structures, by iterative Rayleigh-quotient minimization in a planewave
basis, using a freely available software package \cite{JohnsonJo01}.
Given an arbitrary structure, we can then evaluate the derivative
of the eigenfrequency for a shifting interface, both by the perturbation
eq\@.~(\ref{eq:pert-surf}) and by numerical differentiation of
the eigenfrequencies (here, differentiating a cubic-spline interpolation).%
\begin{figure}
\begin{centering}\includegraphics[width=1\columnwidth]{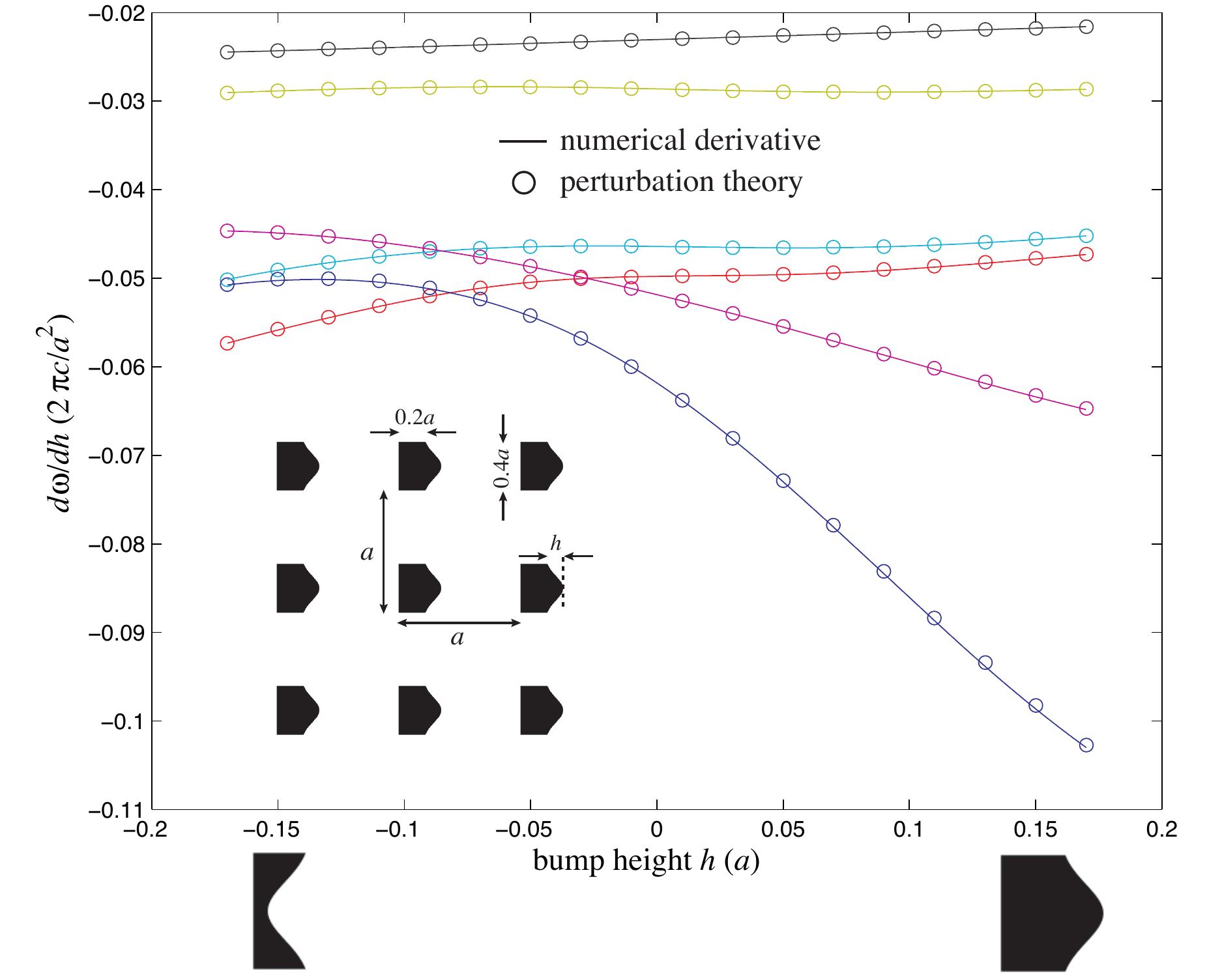}\par\end{centering}

\caption{\label{fig:randbump}Numerical validation of perturbation-theory
formula, applied to compute the derivative $d\omega/dh$ for a Gaussian
{}``bump'' of height $h$ on a square lattice (period $a$) of anisotropic-$\varepsilon$
rectangles (inset) with an eigenfrequency $\omega$ (corresponding
to $\lambda\sim3a$). Positive/negative $h$ indicate bumps/indentations
(see lower-right/left insets for $h=\pm0.15a$), respectively. Solid
lines are numerical differentiation of the eigenfrequency, and dots
are from perturbation theory. The different lines correspond to different
random dielectric tensors $\epstens^{a}$ and $\epstens^{b}$.}
\end{figure}

In particular, we considered a two-dimensional photonic crystal \cite{Joannopoulos95}
consisting of a square lattice (lattice constant $a$) of $0.4a\times0.2a$
dielectric blocks of a material $\epstens^{a}$ surrounded by $\epstens^{b}$,
with Gaussian bumps on one side (inset of Fig\@.~\ref{fig:randbump}).
Here, $\epstens^{a}$ and $\epstens^{b}$ are chosen to be random
symmetric positive-definite matrices with eigenvalues ranging from
2 to 12 for $\epstens^{a}$ and from 1 to 5 for $\epstens^{b}$. On
the right side of each block (along one of the $0.4a$ edges) is a
Gaussian bump of height $h(y)=he^{-y^{2}/2w^{2}}$, with a width $w=0.1a$
and amplitude $h$ (where $h<0$ denotes an indentation). We then
computed the lowest eigenvalue $\omega(A)$ and eigenfields $\vec{E}$
for a set of $h$ values $h/a\in[-0.17,+0.17]$ , at a Bloch wavevector
$\vec{k}=(0,0,0.5)\cdot2\pi/a$ (leading to modes with a vacuum wavelength
$\lambda\sim3a$). Given this data, we then compared the derivative
$d\omega/dh$ as computed by the perturbation equation~(\ref{eq:pert-surf})
compared to the derivative of a cubic-spline fit of the frequency
data. This was repeated for six different random $\epstens^{a}$ and
$\epstens^{b}$. The results, shown in Fig\@.~\ref{fig:randbump},
demonstrate that the perturbation formula indeed predicts the exact
slope $h$ as expected (with tiny discrepancies, due to the finite
resolution, too small to see on this graph).

\section{Application to Sub-Pixel Smoothing}

In any numerical method involving the solution of the full-vector
Maxwell's equations on a discrete grid or its equivalent, such as
the planewave method above \cite{JohnsonJo01} or the finite-difference
time-domain (FDTD) method \cite{Taflove00}, discontinuities in the
dielectric function $\varepsilon$ (and the corresponding field discontinuities)
generally degrade the accuracy of the method, typically reducing it
to only linear convergence with resolution \cite{JohnsonJo01,Ditkowski01}.
Unfortunately, piecewise-continuous $\varepsilon$ is the most common
experimental situation, so a technique to improve the accuracy (without
switching to an entirely different computational method) is desirable.
One simple approach that has been proposed by several authors is to
\emph{smooth} the dielectric function, or equivalently to set the
$\varepsilon$ of each {}``pixel'' to be some average of $\varepsilon$
within the pixel, rather than merely sampling $\varepsilon$ in a
{}``staircase'' fashion \cite{Kaneda97,Dey99,Meade93,Lee99,JohnsonJo01,Nadobny03,Mohammadi05,Farjadpour06}.
Unfortunately, this smoothing itself changes the structure, and therefore
introduces errors. We analyzed this situation in a recent paper for
the FDTD method \cite{Farjadpour06}, and showed that the problem
is closely related to perturbation theory: one desires a smoothing
of $\varepsilon$ that has \emph{zero first-order effect}, to minimize
the error introduced by smoothing and so that the underlying second-order
accuracy can potentially be preserved. At an interface between two
isotropic dielectric materials, the first-order perturbation is given
by eq\@.~(\ref{eq:pert-surf-scalar}), and this leads to an \emph{anisotropic}
smoothing: one averages $\varepsilon^{-1}$ for field components perpendicular
to the interface, and averages $\varepsilon$ for field components
parallel to the interface, a result that had previously been proposed
heuristically by several authors \cite{Meade93,Lee99,JohnsonJo01}.

In this section, we generalize that result to interfaces between anisotropic
materials, and illustrate numerically that it leads to both dramatic
improvements in the absolute magnitude and the convergence rate of
the discretization error. In the anisotropic-interface case, a heuristic
subpixel smoothing scheme was previously proposed \cite{JohnsonJo01},
but we now show that this method was suboptimal: although it is better
than other smoothing schemes, it does not set the first-order perturbation
to zero and therefore does not minimize the error or permit the possibility
of second-order accuracy. Specifically, as discussed more explicitly
below, a second-order smoothing is obtained by averaging $\ptens(\epstens)$
and then inverting $\ptens(\epstens)$ to obtain the smoothed {}``effective''
dielectric tensor. Because this scheme is analytically guaranteed
to eliminate the first-order error otherwise introduced by smoothing,
we expect it to generally lead to the smallest numerical error compared
to competing smoothing schemes, and there is the hope that the overall
convergence rate may be quadratic with resolution.

First, let us analyze how perturbation theory leads to a smoothing
scheme. Suppose that we smooth the underlying dielectric tensor $\epstens(\vec{x})$
into some locally averaged tensor $\bar{\epstens}(\vec{x})$, by some
method to be determined below. This involves a change $\Delta\epstens=\bar{\epstens}-\epstens$,
which is likely to be large near points where $\epstens$ is discontinuous
(and, conversely, is zero well inside regions where $\epstens$ is
constant). In particular, suppose that we employ a smoothing radius
(defined more precisely below) proportional to the spatial resolution
$\Delta x$ of our numerical method, so that $\Delta\epstens$ is
zero {[}or at most $O(\Delta x^{2})]$ except within a distance $\sim\Delta x$
of discontinuous interfaces. To evaluate the effect of this large
perturbation near an interface, we must employ an equivalent reformulation
of eq\@.~(\ref{eq:pert-surf}): \[
\Delta\omega\sim\int\vec{F}^{*}\cdot\Delta\ptens\cdot\vec{F}\, d^{3}\vec{x},\]
where $\Delta\ptens=\ptens(\bar{\epstens})-\ptens(\epstens$). It
is sufficient to look at the perturbation in $\omega$, since (as
we remarked in Sec\@.~\ref{sec:Problem-Formulation}) the same integral
appears in the perturbation theory for many other quantities (such
as scattered power, etc.). If we let $x_{1}$ denote the (local) coordinate
orthogonal to the boundary, then the $x_{1}$ integral is simply proportional
to $\sim\int\Delta\ptens\, dx_{1}+O(\Delta x^{2})$ : since $\vec{F}$
is continuous and $\Delta\ptens=0$ except near the interface, we
can pull $\vec{F}$ out of the $x_{1}$ integral to lowest order.
That means, in order to make the first-order perturbation zero for
all fields $\vec{F}$, it is sufficient to have $\int\Delta\ptens\, dx_{1}=0$.
This is achieved by averaging $\ptens$ as follows.

The most straightforward interpretation of {}``smoothing'' would
be to convolve $\epstens$ with some localized kernel $s(\vec{x})$,
where $\int s(\vec{x})\, d^{3}\vec{x}=1$ and $s(\vec{x})=0$ for
$|\vec{x}|$ greater than some smoothing radius (the support radius)
proportional to the resolution $\sim\Delta x$. That is, $\bar{\epstens}(\vec{x})=\epstens\ast s=\int\epstens(\vec{y})\, s(\vec{x}-\vec{y})\, d^{3}\vec{y}$.
For example, the simplest subpixel smoothing, simply computing the
average of $\epstens$ over each pixel, corresponds to $s=1$ inside
a pixel at the origin and $s=0$ elsewhere. However, this will not
lead to the desired $\int\Delta\ptens=0$ to obtain second order accuracy.
Instead, we employ: \begin{equation}
\bar{\epstens}(\vec{x})=\ptens^{-1}[\ptens(\epstens)\ast s]=\ptens^{-1}\left\{ \int\ptens\left[\epstens(\vec{y})\right]\, s(\vec{x}-\vec{y})\, d^{3}\vec{y}\right\} .\label{eq:eps-average}\end{equation}
where $\ptens^{-1}$ is the inverse of the $\ptens(\epstens)$ mapping,
given by:\begin{equation}
\ptens^{-1}(\ptens)=\left(\begin{array}{ccc}
-\frac{1}{\p_{11}} & -\frac{\p_{12}}{\p_{11}} & -\frac{\p_{13}}{\p_{11}}\\
-\frac{\p_{21}}{\p_{11}} & \quad\p_{22}-\frac{\p_{21}\p_{12}}{\p_{11}}\quad & \p_{23}-\frac{\p_{21}\p_{13}}{\p_{11}}\\
-\frac{\p_{31}}{\p_{11}} & \p_{32}-\frac{\p_{31}\p_{12}}{\p_{11}} & \p_{33}-\frac{\p_{31}\p_{13}}{\p_{11}}\end{array}\right).\label{eq:ptens-inverse}\end{equation}
The reason why eq\@.~(\ref{eq:eps-average}) works, regardless of
the smoothing kernel $s(\vec{x})$, is that \begin{align}
\int\Delta\ptens\, d^{3}\vec{x} & =\int d^{3}\vec{x}\left\{ \int\ptens\left[\epstens(\vec{y})\right]\, s(\vec{x}-\vec{y})\, d^{3}\vec{y}-\epstens(\vec{x})\right\} \nonumber \\
 & =\int d^{3}\vec{y\,}\ptens\left[\epstens(\vec{y})\right]\left\{ \int\, s(\vec{x}-\vec{y})\, d^{3}\vec{y}-1\right\} \nonumber \\
 & =0.\label{eq:convolution-delta-tau-integral}\end{align}
 This guarantees that the integral of $\Delta\ptens$ is zero over
all space, but above we required what appears to be a stronger condition,
that the local, interface-perpendicular integral $\int\Delta\ptens\, dx_{1}$
be zero (at least to first order). However, in a small region where
the interface is locally flat (to first order in the smoothing radius),
$\Delta\ptens$ must be a function of $x_{1}$ only by translational
symmetry, and therefore (\ref{eq:convolution-delta-tau-integral})
implies that $\int\Delta\ptens\, dx_{1}=0$ by itself. Although the
above convolution formulas may look complicated, for the simplest
smoothing kernel $s(\vec{x})$ the procedure is quite simple: in each
pixel, average $\ptens(\epstens)$ in the pixel and then apply $\ptens^{-1}$
to the result. (This is not any more difficult to apply than the procedure
implemented in Ref\@.~\cite{JohnsonJo01}, for example.)

Strictly speaking, the use of this smoothing does not guarantee second-order
accuracy, even if the underlying numerical method is nominally second-order
accurate or better. For one thing, although we have canceled the first-order
error due to smoothing, it may be that the next-order correction is
not second-order. Precisely this situation occurs if one has a structure
with sharp dielectric corners, edges, or cusps, as discussed in Ref\@.~\cite{JohnsonIb02}:
in this case, smoothing leads to a convergence rate between first
order (what would be obtained with no smoothing) and second order,
with the exponent determined by the nature of the field singularity
that occurs at the corner.

\subsection{Numerical smoothing validation}

\begin{figure}
\begin{centering}\includegraphics[width=1\columnwidth]{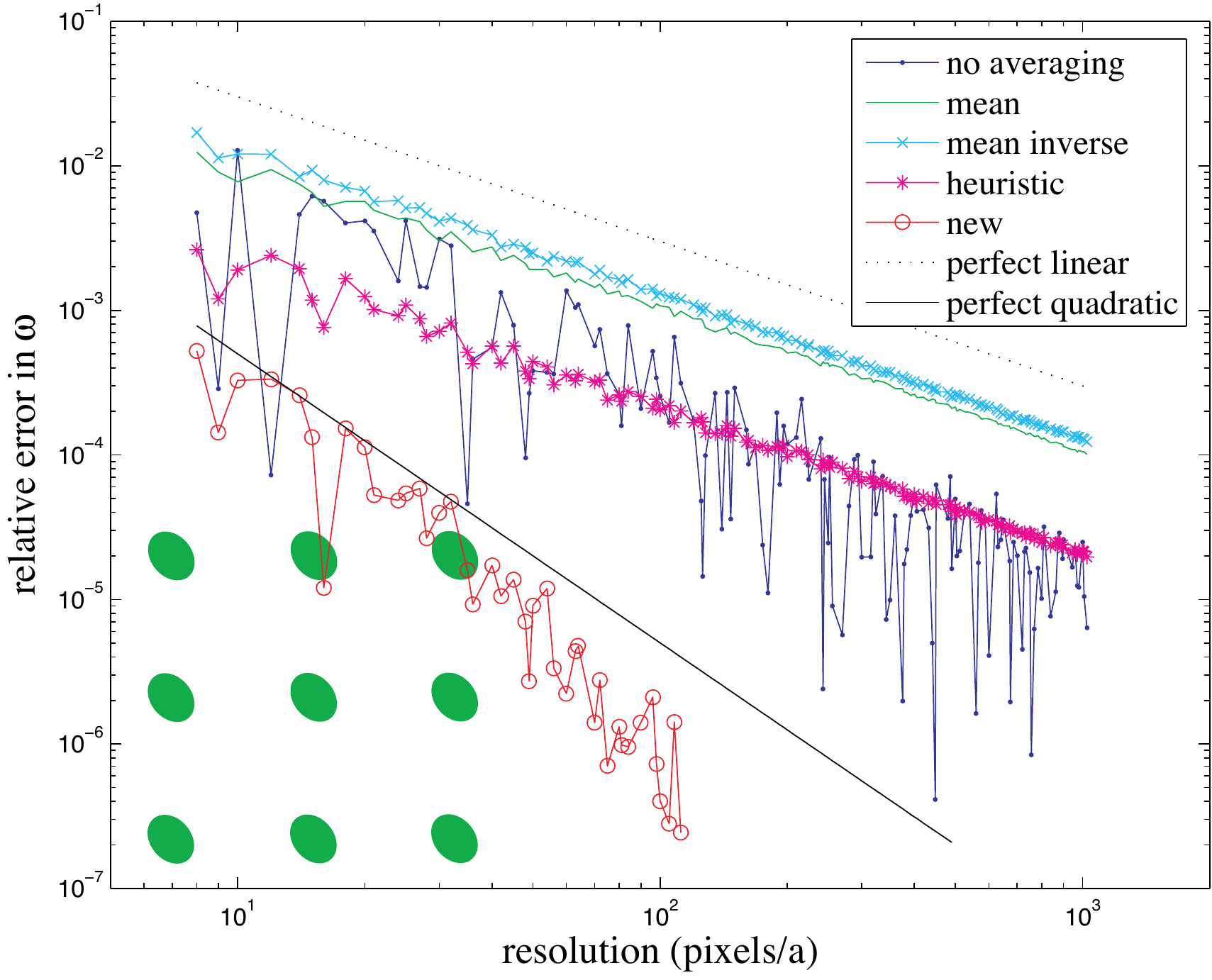}\par\end{centering}

\caption{\label{fig:ellipse3}Relative error $\Delta\omega/\omega$ for an
eigenmode calculation with a square lattice (period $a$) of 2d anisotropic
ellipses (green inset), versus spatial resolution, for a variety of
sub-pixel smoothing techniques. Straight lines for perfect linear
(black dashed) and perfect quadratic (black solid) convergence are
shown for reference.}
\end{figure}

As a simple illustration of the efficacy of the subpixel smoothing
we propose in eq\@.~(\ref{eq:eps-average}), let us consider a two-dimensional
example problem: a square lattice (period $a$) of ellipses made of
$\epstens^{a}$ surrounded by $\epstens^{b}$, where we will find
the lowest-$\omega$ Bloch eigenmode. As above, we choose the dielectric
tensors to be random positive-definite symmetric matrices with random
eigenvalues in $[2,12]$ for $\epstens^{a}$ and in $[1,5]$ for $\epstens^{b}$,
and the ellipses are oriented at an arbitrary angle, at an arbitrary
Bloch wavevector $\vec{k}a/2\pi=(0.1,0.2,0.3)$, to avoid fortuitous
symmetry effects. (The vacuum wavelength $\lambda$ corresponding
to the eigenfrequency $\omega$ is $\lambda=5.03a$.) For each resolution
$\Delta x$, we assign an $\bar{\epstens}$ to each pixel by computing
$\ptens^{-1}$ of the average of $\ptens(\epstens)$ within that pixel.
Then, we compute the relative error $\Delta\omega/\omega$ (compared
to a calculation at a much higher resolution) as a function of resolution.
For comparison, we also consider four other smoothing techniques:
no smoothing, averaging $\epstens$ in each pixel \cite{Dey99}, averaging
$\epstens^{-1}$ in each pixel, and a heuristic anisotropic averaging
proposed by Ref\@.~\cite{JohnsonJo01} in analogy to the scalar
case. The results are shown in Fig\@.~\ref{fig:ellipse3}, based
on the same planewave method as above \cite{JohnsonJo01}, and show
that the new smoothing technique clearly leads to the lowest errors
$\Delta\omega/\omega$. Also, whereas the other methods yield clearly
first-order convergence, the new method seems to exhibit roughly second-order
convergence. The no-smoothing case has extremely erratic errors, as
is typical for stair-casing phenomena.%

\begin{figure}
\begin{centering}\includegraphics[width=1\columnwidth]{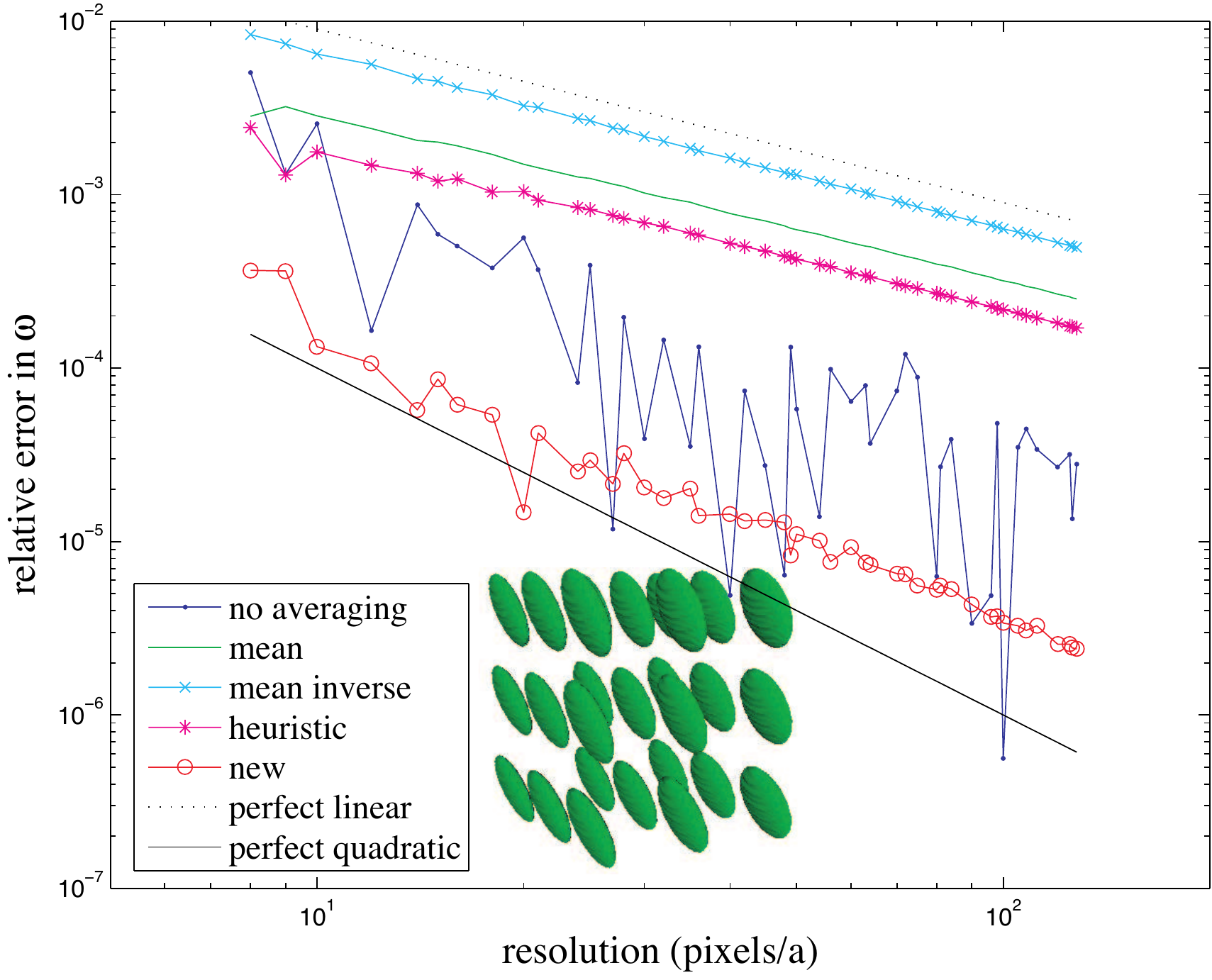}\par\end{centering}

\caption{\label{fig:3Dellipse3}Relative error $\Delta\omega/\omega$ for
an eigenmode calculation with cubic lattice (period $a$) of 3d anisotropic
ellipsoids (green inset), versus spatial resolution, for a variety
of sub-pixel smoothing techniques. Straight lines for perfect linear
(black dashed) and perfect quadratic (black solid) convergence are
shown for reference.}
\end{figure}

In Fig\@.~\ref{fig:3Dellipse3}, we also show results from a similar
calculation in three dimensions. Here, we look at the lowest eigenmode
of a cubic lattice (period $a$) of 3d ellipsoids (oriented at a random
angle) made of $\epstens^{a}$ surrounded by $\epstens^{b}$, both
random positive-definite symmetric matrices as above. The frequency
$\omega$, at an arbitrarily chosen wavevector $\vec{k}a/2\pi=(0.4,0.3,0.1)$,
corresponds to a vacuum wavelength $\lambda=3.14a$. Again, the new
method almost always has the lowest error by a wide margin, especially
if the unpredictable dips of the no-smoothing case are excluded, and
is the only one to exhibit (apparently) better than linear convergence.%

Our previous heuristic proposal from Ref\@.~\cite{JohnsonJo01},
while better than the other smoothing schemes (and less erratic than
no smoothing), is clearly inferior to the new method. Previously,
we had observed what seemed to have been quadratic convergence from
the heuristic scheme \cite{JohnsonJo01}, but this result seems to
have been fortuitous---as we demonstrated recently, even non-second-order
schemes can sometimes appear to have second-order convergence over
some range of resolutions for a particular geometry \cite{Farjadpour06}.
The key distinction of the new scheme, that lends us greater confidence
in it than one or two examples can convey, is that it is no longer
heuristic. The new smoothing scheme is based on a clear analytical
criterion---setting the first-order perturbative effect of the smoothing
to zero---that explains why it should be an accurate choice in a wide
variety of circumstances.

\section{Concluding Remarks}

We have shown how to correctly treat lowest-order perturbations to
a boundary between two anisotropic materials, a problem for which
previous approaches had been stymied by the complicated discontinuous
boundary conditions on the electric field. This result immediately
led to an improved subpixel smoothing scheme for discretized numerical
methods---we demonstrated it for a planewave method, but we expect
that it will similarly be applicable to other methods, e.g. FDTD \cite{Farjadpour06}.
The same result can also be applied to constructing an effective-medium
theory for subwavelength multilayer films of anisotropic materials.
Moreover, in the process of deriving our perturbative result, we developed
a \emph{local coordinate-transform} approach that may be useful in
treating many other types of interface perturbations, because it circumvents
the difficulty of shifting discontinuities without requiring one to
construct an explicit coordinate transformation.

\begin{acknowledgments}
This work was supported in part by Dr. Dennis Healy of DARPA MTO,
under award N00014-05-1-0700 administered by the Office of Naval Research.
We are also grateful to I.~Singer at MIT for helpful discussions.
\end{acknowledgments}

\section*{Appendix: Coordinate transformation of Maxwell's equations}

As discussed in Sec\@.~\ref{sub:Coordinate-perturbations} above,
any differentiable coordinate transformation of Maxwell's equations
can be recast as merely a transformation of $\varepsilon$ and $\mu$,
with the same solutions $\vec{E}$ and $\vec{H}$ only multiplied
by a matrix in addition to the coordinate change \cite{Ward96}. This
result has been exploited by Pendry et~al. to obtain a number of
beautiful analytical results from cylindrical superlenses \cite{Pendry03}
to {}``invisibility'' cloaks \cite{Pendry06}. It (and related ideas)
can be used to derive coupled-mode expressions for bending loss in
optical waveguides \cite{Katsenelenbaum98,Johnson01:og}. A similar
result has also been employed to design perfectly-matched layers (PML),
via a \emph{complex} coordinate stretching, to truncate numerical
grids \cite{Teixeira98}. It is likely that there are many other applications,
as well as equivalent derivations, that we are not aware of. Here,
we review the proof in a compact form, generalized to arbitrary anisotropic
media. (Most previous derivations seem to have been for isotropic
media in at least one coordinate frame \cite{Ward96}, or for coordinate
transformations with purely diagonal Jacobians $\Jactens$ where $\Jac_{ii}$
depends only on $x_{i}$ \cite{Teixeira98}, or for constant affine
coordinate transforms \cite{Lindell92}.)

We begin with the usual Maxwell's equations for Euclidean space (in
natural units): \begin{eqnarray}
\nabla\times\vec{H} & = & \epstens\cdot\frac{\partial\vec{E}}{\partial t}+\vec{J}\label{eq:maxwell-H}\\
\nabla\times\vec{E} & = & -\mutens\cdot\frac{\partial\vec{H}}{\partial t}\label{eq:maxwell-E}\\
\nabla\cdot(\epstens\cdot\vec{E}) & = & \rho\label{eq:maxwell-Gauss-E}\\
\nabla\cdot(\mutens\cdot\vec{H}) & = & 0,\label{eq:maxwell-Gauss-H}\end{eqnarray}
where $\vec{J}$ and $\rho$ are the usual free current and charge
densities, respectively. We will proceed in index notation, employing
the Einstein convention whereby repeated indices are summed over.
Ampere's Law, eq\@.~(\ref{eq:maxwell-H}), is now expressed:\begin{equation}
\partial_{a}H_{b}\epsilon_{abc}=\varepsilon_{cd}\frac{\partial E_{d}}{\partial t}+J_{c}\label{eq:maxwell-H-Einstein}\end{equation}
 where $\epsilon_{abc}$ is the usual Levi-Civita permutation tensor
and $\partial_{a}=\partial/\partial x_{a}$. Under a coordinate change
$\vec{x}\mapsto\vec{x}'$, if we let $\Jac_{ab}=\frac{\partial x_{a}'}{\partial x_{b}'}$
be the (non-singular) Jacobian matrix associated with the coordinate
transform (which may be a function of $\vec{x}$), we have \begin{equation}
\partial_{a}=\Jac_{ba}\partial_{b}'.\label{eq:del-Jac}\end{equation}
Furthermore, as in eqs\@.~(\ref{eq:E-trans}--\ref{eq:H-trans}),
let \begin{eqnarray}
E_{a} & = & \Jac_{ba}E_{b}',\label{eq:E-trans-Einstein}\\
H_{a} & = & \Jac_{ba}H_{b}'.\label{eq:H-trans-Einstein}\end{eqnarray}
Hence, eq\@.~(\ref{eq:maxwell-H-Einstein}) becomes \begin{equation}
\Jac_{ia}\partial_{i}'\Jac_{jb}H'_{j}\epsilon_{abc}=\varepsilon_{cd}\Jac_{ld}\frac{\partial E'_{l}}{\partial t}+J_{c}.\label{eq:maxwell-H-transform-1}\end{equation}
 Here, the $\Jac_{ia}\partial_{i}'=\partial_{a}$ derivative falls
on both the $\Jac_{jb}$ and $H'_{j}$ terms, but we can eliminate
the former thanks to the $\epsilon_{abc}$: $\partial_{a}\Jac_{jb}\epsilon_{abc}=0$
because $\partial_{a}\Jac_{jb}=\partial_{b}\Jac_{ja}$. Then, again
multiplying both sides by the Jacobian $\Jac_{kc}$, we obtain \begin{equation}
\Jac_{kc}\Jac_{jb}\Jac_{ia}\partial_{i}'H'_{j}\epsilon_{abc}=\Jac_{kc}\varepsilon_{cd}\Jac_{ld}\frac{\partial E'_{l}}{\partial t}+\Jac_{kc}J_{c}\label{eq:maxwell-H-transform-2}\end{equation}
 Noting that $\Jac_{ia}\Jac_{jb}\Jac_{kc}\epsilon_{abc}=\epsilon_{ijk}\det\Jactens$
by definition of the determinant, we finally have\begin{equation}
\partial_{i}'H'_{j}\epsilon_{ijk}=\frac{1}{\det\Jactens}\Jac_{kc}\varepsilon_{cd}\Jac_{ld}\frac{\partial E'_{l}}{\partial t}+\frac{\Jac_{kc}J_{c}}{\det\Jactens}\label{eq:maxwell-H-transform-3}\end{equation}
 or, back in vector notation, \begin{equation}
\nabla'\times\vec{H}'=\frac{\Jactens\cdot\epstens\cdot\Jactens^{T}}{\det\Jactens}\cdot\frac{\partial\vec{E}'}{\partial t}+\vec{J}',\label{eq:maxwell-H-transform}\end{equation}
where $\vec{J}'=\Jactens\cdot\vec{J}/\det\Jactens$. Thus, we see
that we can interpret Ampere's Law in arbitrary coordinates as the
usual equation in Euclidean coordinates, as long as we replace the
materials etc. by eqs\@.~(\ref{eq:E-trans}--\ref{eq:eps-trans}).
By an identical argument, we obtain\begin{equation}
\nabla'\times\vec{E}'=-\frac{\Jactens\cdot\mutens\cdot\Jactens^{T}}{\det\Jactens}\cdot\frac{\partial\vec{H}'}{\partial t},\label{eq:maxwell-E-transform}\end{equation}
which yields the corresponding transformation (\ref{eq:mu-trans})
for $\mutens$.

The transformation of the remaining divergence equations into equivalent
forms in the new coordinates is also straightforward. Gauss's Law,
eq\@.~(\ref{eq:maxwell-Gauss-E}), becomes \begin{align}
\rho & =\partial_{a}\varepsilon_{ab}E_{b}=\Jac_{ia}\partial_{i}'\varepsilon_{ab}\Jac_{jb}E_{j}'=\Jac_{ia}\partial_{i}'(\det\Jactens)\Jac_{ak}^{-1}\varepsilon_{kj}'E_{j}'\nonumber \\
 & =(\det\Jactens)\partial_{i}'\varepsilon_{ij}'E_{j}'+(\partial_{a}\Jac_{ak}^{-1}\det\Jactens)\varepsilon_{kj}'E_{j}'\nonumber \\
 & =(\det\Jactens)\partial_{i}'\varepsilon_{ij}'E_{j}',\label{eq:maxwell-Gauss-E-transform}\end{align}
which gives $\nabla'\cdot(\epstens'\cdot\vec{E}')=\rho'$ for $\rho'=\rho/\det\Jactens$.
Similarly for eq\@.~(\ref{eq:maxwell-Gauss-H}). Here, we have used
the fact that\begin{equation}
\partial_{a}\Jac_{ak}^{-1}\det\Jactens=\partial_{a}\epsilon_{anm}\epsilon_{kij}\Jac_{in}\Jac_{jm}/2=0,\label{eq:Jac-inv-deriv-identity}\end{equation}
from the cofactor formula for the matrix inverse, and recalling that
$\partial_{a}\Jac_{jb}\epsilon_{abc}=0$ from above. In particular,
note that $\rho=0\Longleftrightarrow\rho'=0$ and $\vec{J}=0\Longleftrightarrow\vec{J}'=0$,
so a non-singular coordinate transformation preserves the absence
(or presence) of sources.


\begin{thebibliography}{10}

\bibitem{JohnsonIb02}
S.~G. Johnson, M.~Ibanescu, M.~A. Skorobogatiy, O.~Weisberg, J.~D.
  Joannopoulos, and Y.~Fink, ``Perturbation theory for {Maxwell}'s equations
  with shifting material boundaries,'' {\em Physical Review~E}, vol.~65,
  p.~066611, 2002.

\bibitem{Smith04}
D.~R. Smith, J.~B. Pendry, and M.~C.~K. Wiltshire, ``Metamaterials and negative
  refractive index,'' {\em Science}, vol.~305, pp.~788--792, 2004.

\bibitem{Pendry03}
J.~Pendry, ``Perfect cylindrical lenses,'' {\em Optics Express}, vol.~11,
  pp.~755--760, 2003.

\bibitem{Pendry06}
J.~B. Pendry, D.~Schurig, and D.~R. Smith, ``Controlling electromagnetic
  fields,'' {\em Science}, vol.~312, pp.~1780--1782, 2006.

\bibitem{Farjadpour06}
A.~Farjadpour, D.~Roundy, A.~Rodriguez, M.~Ibanescu, P.~Bermel, J.~D.
  Joannopoulos, S.~G. Johnson, and G.~W. Burr, ``Improving accuracy by subpixel
  smoothing in the finite-difference time domain,'' {\em Optics Letters},
  vol.~31, pp.~2972--2974, 2006.

\bibitem{JohnsonJo01}
S.~Johnson and J.~Joannopoulos, ``Block-iterative frequency-domain methods for
  {Maxwell}'s equations in a planewave basis,'' {\em Optics Express}, vol.~8,
  pp.~173--190, 2001.

\bibitem{Ditkowski01}
A.~Ditkowski, K.~Dridi, and J.~S. Hesthaven, ``Convergent cartesian grid
  methods for {Maxwell}'s equations in complex geometries,'' {\em J.~Comp.
  Phys.}, vol.~170, pp.~39--80, 2001.

\bibitem{Hill81}
N.~R. Hill, ``Integral-equation perturbative approach to optical scattering
  from rough surfaces,'' {\em Physical Review~B}, vol.~24, no.~12,
  pp.~7112--7120, 1981.

\bibitem{Lohmeyer99}
M.~Lohmeyer, N.~Bahlmann, and P.~Hertel, ``Geometry tolerance estimation for
  rectangular dielectric waveguide devices by means of perturbation theory,''
  {\em Optics Communications}, vol.~163, pp.~86--94, 1999.

\bibitem{Skorobogatiy02:curvi}
M.~Skorobogatiy, S.~A. Jacobs, S.~G. Johnson, and Y.~Fink, ``Geometric
  variations in high index-contrast waveguides, coupled mode theory in
  curvilinear coordinates,'' {\em Optics Express}, vol.~10, pp.~1227--1243,
  2002.

\bibitem{Skorobogatiy04}
M.~Skorobogatiy, ``Modeling the impact of imperfections in high-index-contrast
  photonic waveguides,'' {\em Physical Review~E}, vol.~70, p.~046609, 2004.

\bibitem{Johnson05:bump}
S.~G. Johnson, M.~L. Povinelli, M.~Solja{\v{c}}i{\'{c}}, A.~Karalis, S.~Jacobs,
  and J.~D. Joannopoulos, ``Roughness losses and volume-current methods in
  photonic-crystal waveguides,'' {\em Appl. Phys. B}, vol.~81, pp.~283--293,
  2005.

\bibitem{Meade93}
R.~D. Meade, A.~M. Rappe, K.~D. Brommer, J.~D. Joannopoulos, and O.~L.
  Alerhand, ``Accurate theoretical analysis of photonic band-gap materials,''
  {\em Physical Review~B}, vol.~48, pp.~8434--8437, 1993.
\newblock Erratum: S.~G. Johnson, {\em ibid.} {\bf 55}, 15942 (1997).

\bibitem{Ward96}
A.~J. Ward and J.~B. Pendry, ``Refraction and geometry in {Maxwell}'s
  equations,'' {\em Journal of Modern Optics}, vol.~43, no.~4, pp.~773--793,
  1996.

\bibitem{Tannoudji77}
C.~Cohen-Tannoudji, B.~Din, and F.~Lalo{\"{e}}, {\em Quantum Mechanics}.
\newblock Paris: Hermann, 1977.

\bibitem{Joannopoulos95}
J.~D. Joannopoulos, R.~D. Meade, and J.~N. Winn, {\em Photonic Crystals:
  Molding the Flow of Light}.
\newblock Princeton Univ. Press, 1995.

\bibitem{Johnson01:og}
S.~G. Johnson, M.~Ibanescu, M.~Skorobogatiy, O.~Weisberg, T.~D. Engeness,
  M.~Solja{\v{c}}i{\'{c}}, S.~A. Jacobs, J.~D. Joannopoulos, and Y.~Fink,
  ``Low-loss asymptotically single-mode propagation in large-core {O}mni{G}uide
  fibers,'' {\em Optics Express}, vol.~9, no.~13, pp.~748--779, 2001.

\bibitem{Marcuse74}
D.~Marcuse, {\em Theory of Dielectric Optical Waveguides}.
\newblock San Diego: Academic Press, second~ed., 1991.

\bibitem{Katsenelenbaum98}
B.~Z. Katsenelenbaum, L.~Mercader~del R{\'{i}}o, M.~Pereyaslavets,
  M.~Sorolla~Ayza, and M.~Thumm, {\em Theory of Nonuniform Waveguides: The
  Cross-Section Method}.
\newblock London: Inst. of Electrical Engineers, 1998.

\bibitem{Johnson02:adiabatic}
S.~G. Johnson, P.~Bienstman, M.~Skorobogatiy, M.~Ibanescu, E.~Lidorikis, and
  J.~D. Joannopoulos, ``Adiabatic theorem and continuous coupled-mode theory
  for efficient taper transitions in photonic crystals,'' {\em Physical
  Review~E}, vol.~66, p.~066608, 2002.

\bibitem{Snyder83}
A.~W. Snyder and J.~D. Love, {\em Optical Waveguide Theory}.
\newblock London: Chapman and Hall, 1983.

\bibitem{Kuznetsov83}
M.~Kuznetsov and H.~A. Haus, ``Radiation loss in dielectric waveguide
  structures by the volume current method,'' {\em IEEE Journal of Quantum
  Electronics}, vol.~19, pp.~1505--1514, 1983.

\bibitem{Chew95}
W.~C. Chew, {\em Waves and Fields in Inhomogeneous Media}.
\newblock New York, NY: IEEE Press, 1995.

\bibitem{LeungLi94}
P.~T. Leung, S.~Y. Liu, and K.~Young, ``Completeness and time-independent
  perturbation of the quasinormal modes of an absorptive and leaky cavity,''
  {\em Physical Review~A}, vol.~49, pp.~3982--3989, 1994.

\bibitem{Munkres75}
J.~R. Munkres, {\em Topology: A First Course}.
\newblock Englewood Cliffs, NJ: Prentice-Hall, 1975.

\bibitem{Taflove00}
A.~Taflove and S.~C. Hagness, {\em Computational Electrodynamics: The
  Finite-Difference Time-Domain Method}.
\newblock Norwood, MA: Artech, 2000.

\bibitem{Kaneda97}
N.~Kaneda, B.~Houshmand, and T.~Itoh, ``{FDTD} analysis of dielectric
  resonators with curved surfaces,'' {\em IEEE Transactions on Microwave Theory
  and Techniques}, vol.~45, no.~9, pp.~1645--1649, 1997.

\bibitem{Dey99}
S.~Dey and R.~Mittra, ``A conformal finite-difference time-domain technique for
  modelling cylindrical dielectric resonators,'' {\em IEEE Transactions on
  Microwave Theory and Techniques}, vol.~47, no.~9, pp.~1737--1739, 1999.

\bibitem{Lee99}
J.-Y. Lee and N.-H. Myung, ``Locally tensor conformal {FDTD} method for
  modeling arbitrary dielectric surfaces,'' {\em Microwave and Optical Tech.
  Lett.}, vol.~23, no.~4, pp.~245--249, 1999.

\bibitem{Nadobny03}
J.~Nadobny, D.~Sullivan, W.~Wlodarczyk, P.~Deuflhard, and P.~Wust, ``A 3-d
  tensor {FDTD}-formulation for treatment of sloped interfaces in electrically
  inhomogeneous media,'' {\em IEEE Transactions on Antennas and Propagation},
  vol.~51, no.~8, pp.~1760--1770, 2003.

\bibitem{Mohammadi05}
A.~Mohammadi, H.~Nadgaran, and M.~Agio, ``Contour-path effective permittivities
  for the two-dimensional finite-difference time-domain method,'' {\em Optics
  Express}, vol.~13, no.~25, pp.~10367--10381, 2005.

\bibitem{Teixeira98}
F.~L. Teixeira and W.~C. Chew, ``General closed-form {PML} constitutive tensors
  to match arbitrary bianisotropic and dispersive linear media,'' {\em IEEE
  Microwave and Guided Wave Letters}, vol.~8, no.~6, pp.~223--225, 1998.

\bibitem{Lindell92}
I.~V. Lindell, {\em Methods for Electromagnetic Fields Analysis}.
\newblock Oxford, U.K.: Oxford Univ. Press, 1992.

\end{thebibliography}
\end{document}